# Multi-energy CT reconstruction using nonlocal total nuclear generalized variation

Cheng Kai, Jiang Min, Jianqiao Yu, Sun Yi

*School of Information and Communication, Dalian University of technology,Dalian,*

*Liaoning* 116024,*China*

**Abstract:** Multi-energy CT based on compression sensing theory with sparse-view sampling can effectively reduce radiation dose and maintain the quality of the reconstructed image. However, when the projection data are noisy, the reconstructed image can be still seriously degraded. In order to address this problem, we extend the total nuclear variation(TNV)regularization and propose the nonlocal total nuclear generalized variation (NLTNGV) regularization term. NLTNGV is constructed by using the low rank property of both the nonlocal Jacobian matrix and the local second derivative of the image at different energy spectrum, which can be seen as a more robust structure similarity measure. By employing NLTNGV, the proposed reconstruction method can effectively utilize the sparsity of nonlocal gradients and local second order derivatives and structural similarity of multi-energy CT images to effectively recover the image structure features in noisy case. The experimental results show that the proposed method can enhance the noise-resistant performance and preserve the fine details of the reconstructed images in comparison with TNV-based method.

**Key words:** Multi-energy CT image reconstruction; nonlocal total nuclear generalized variation(NLTNGV) ; dual low rank constraints; nonlocal self similarity; convex optimization model

## 1. Introduction

Compared with single energy CT, multi-energy CT can provide the information of the reconstructed image along energy spectrum dimension, which can effectively determine composition of material components [1]. However, multi-energy CT needs to obtain projection data of different energy spectrum, thus may increase the risk of radiation dose. Sparse view sampling can efficiently reduce the radiation dose, but the reconstructed images will suffer from striking artifacts due to insufficient projection data. As the CT images at different energy spectrum characterize the structure information of the same object, they share structural similarity. Exploiting such prior information, also combined with the sparsity prior information in transform domain of the reconstructed image, the insufficient projection data problem can be overcome. Gao et al. [2] proposed a multi-energy CT reconstruction algorithm based on low rank and sparse decomposition (PRISM) by using the robust principal component analysis to use the above prior information, which effectively improved the reconstruction quality compared with the traditional total variation (TV) regularization method. Kim et al. [3] proposed a multi-energy CT image reconstruction algorithm based on the low rank of local image patches, which further improved the reconstruction quality compared with PRISM method. In [4], the total nuclear variation (TNV) minimization based multi-energy CT reconstruction method was proposed. This method uses the prior information that the gradient vectors of pixels at the same spatial location of the reconstructed image are approximately parallel, but it tends to smooth out the details of the image in large noise case. Some scholars also proposed dictionary learning based reconstruction method for multi-energy CT. For example, [5] and [6] proposed tensor dictionary, image gradient l0-norm and tensor dictionary for multi-energy CT reconstruction. Compared with the fixed sparse transformation, better reconstruction results can be obtained. However, it needs to pre-train the dictionary or learn the dictionary in the process of iterative optimization, leading to higher computational complexity. In the process of CT imaging, noise is an important degradation factor for the reconstructed CT images especially for low dose CT in medical imaging or micro-CT with limited power in industrial application. Considering the nonlocal self similarity prior information of the reconstructed image, the quality of reconstructed image can be significantly improved with noisy projection data. For example, [7]-[11] proposed reconstruction method based on nonlocal means, nonlocal total variation (nonlocal TV), nonlocal spectral similarity based model, BM3D with low rank constraint and nonlocal tensor decomposition, respectively.

Aiming at reducing the striking artifacts caused by insufficient data due to sparse view sampling and the artifacts caused by the noise interference, we generalize the traditional TNV regularization and propose NLTNGV regularization. The NLTNGV can be regarded as a more robust structure similarity measure which is based on nonlocal gradient (graph derivative) and local second order derivatives. The NLTNGV regularization can not only promote the sparsity of nonlocal gradients and local second order derivatives of the reconstructed image at each energy spectrum, but also can promote the nonlocal gradient local second order derivative of the image at each energy spectrum to be approximately parallel, respectively. Moreover, NLTNGV has the advantage of being convex.

By using the favorable characteristics of NLTNGV regularization, a multi-energy CT reconstruction method for sparse noisy projection data is proposed in this work. The proposed method can not only effectively couple the structure information of each energy spectrum image, but also effectively utilize the nonlocal self similarity redundant prior information to effectively improve the noise suppression performance and restore the details of the image, thus improve the overall quality of reconstruction. In addition, the proposed method only needs to solve a convex optimization model, hence it is guaranteed to converge to the global optimal solution.

**2. Method**

The sparsity of gradients, the nonlocal self similarity in spatial domain and structural similarity are the most important prior information for multi-energy CT images. In order to take advantage of the above priori information, we first define NLTNGV regularization term, then introduce it into the multi-energy CT image reconstruction model to effectively improve the robustness against noise and restore the fine details of the reconstructed image.

**2.1 Multi-energy CT forward projection model**

At present, there are several ways to acquire multi-energy projection data, such as two sources and two detectors, single source and "double-layer" detector, single source and photon counting detector, voltage switching and so on [12]. Among them, multi-energy spectrum CT based on voltage switching manner has the advantage of no need to change the hardware of the existing CT system. As shown in Figure 1, the voltage of X-ray source ($E_1, E_2, \cdots, E_M, E_1, E_2 \cdots, E_M, \cdots$) is switched in turn during the scanning process to acquire the projection data of M different energy spectrum. In practical application, dual energy or three energy is most commonly used, so M is taken as 2 or 3 in general.

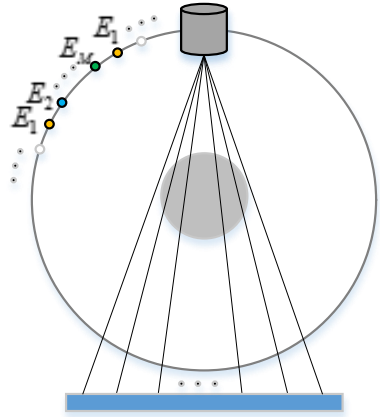

**Figure 1** Voltage switching scanning manner for multi-energy CT

The forward projection model of multi-energy CT in the above scanning manner can be expressed as

$$\begin{aligned} A_1 \mathbf{u}_1 &= \mathbf{p}_1 \\ A_2 \mathbf{u}_2 &= \mathbf{p}_2 \\ &\vdots \\ A_m \mathbf{u}_m &= \mathbf{p}_m \\ &\vdots \\ A_M \mathbf{u}_M &= \mathbf{p}_M \end{aligned} \quad (1)$$

Where $A_m \in R^{N \times N_1 N_2}$, $\mathbf{p}_m \in R^{N \times 1}$, $\mathbf{u}_m \in R^{N_1 N_2 \times 1}$ represents the projection matrix and projection data and the vectorization of the reconstructed image $\mathbf{U}_m \in R^{N_1 \times N_2}$ of the $m^{th}$ energy spectrum respectively; $N = n_d \times S$ is the total number of projection data, $n_d$ is the number of detector elements and $S$ is the number of sampling views at each energy spectrum; $N_1, N_2$ is the number of rows and columns of $\mathbf{U}_m$. For simplification, equation (1) is denoted as $A\mathbf{u} = \mathbf{p}$.

**2.2 NLTNGV regularization**

For multi-energy CT images, although the X-ray attenuation characteristics of materials varies with energy, resulting in different values and contrast of reconstructed images at different energy, the structures of these images are very similar. In order to establish the correlation of image gradients at different energy spectrum to utilize such prior information, the gradient vector of single channel image is extended to Jacobian matrix of multi-channel image in [4]. For pixel $\mathbf{U}(i,j)$, the Jacobian matrix is defined as

$$\mathbf{JU}(i,j) = \begin{pmatrix} \nabla_x \mathbf{U}_1(i,j) & \nabla_y \mathbf{U}_1(i,j) \\ \vdots & \vdots \\ \nabla_x \mathbf{U}_m(i,j) & \nabla_y \mathbf{U}_m(i,j) \\ \vdots & \vdots \\ \nabla_x \mathbf{U}_M(i,j) & \nabla_y \mathbf{U}_M(i,j) \end{pmatrix} \qquad (2)$$

Where $i,j$ represents the row and column index of the image respectively. $\nabla_x, \nabla_y$ are the first-order finite difference along the horizontal direction and the vertical direction respectively, defined as

$$\nabla_x \mathbf{U}_m(i,j) = \begin{cases} \mathbf{U}_m(i+1,j) - \mathbf{U}_m(i,j), 0 < i < N_1 \\ 0, i = N_1 \end{cases} \qquad (3)$$

$$\nabla_y \mathbf{U}_m(i,j) = \begin{cases} \mathbf{U}_m(i,j+1) - \mathbf{U}_m(i,j), 0 < j < N_2 \\ 0, j = N_2 \end{cases} \qquad (4)$$

Therefore, $\mathbf{JU}(i,j) \in \mathbf{R}^{M \times 2}$ characterizes gradient information of the pixels at the location $(i,j)$ with all energy spectrum.

In spite of the efficiency that TNV has demonstrated, it may still suffer from some limitations. One limitation of TNV is that it uses standard gradient vector to couple the structural similarity of images at different energy spectrum, which only involves the horizontal and vertical neighbor information of a pixel, thus not very robust to noise. In addition, TNV only uses the first derivative information and ignores the higher derivative information, which tends to suffer from stepwise effect in the smooth region of the image.

In order to effectively utilize the redundant prior information of nonlocal self similarity to improve the noise-resistant performance and better couple the structural features of multi-energy CT image, we construct a more robust structural similarity measure with respect to first-order derivative by using the graph derivative [13] instead of standard gradient vector. Then we define the structural similarity measure with respect to the higher order derivatives, which can effectively alleviate the stepwise effect of TNV.

For vectorized image $\mathbf{u} \in R^{N_1 N_2 \times 1}$, the graph derivatives of pixel $\mathbf{u}(n)$ ($1 \leq n \leq N_1 \times N_2$) is defined as

$$\nabla_{w,q}\mathbf{u}(n) = (\mathbf{u}(n) - \mathbf{u}(q))\sqrt{w(n,q)}, \forall q \in \Omega_n, q_{\min} \leq q \leq q_{\max} \tag{5}$$

Where $q$ represents the pixel index in the current search window $\Omega_n$ which is centered at pixel $\mathbf{u}(n)$ with size $n_1 \times n_1$. $q_{\min}, q_{\max}$ are the minimum and maximum values that can be taken for $q$ respectively. The weight $w(n,q)$ measures the similarity of pixel pairs $\mathbf{u}(n)$ and $\mathbf{u}(q)$, defined as

$$w(n,q) = \exp(-\frac{\|\mathbf{b}(n) - \mathbf{b}(q)\|}{2h^2}) \tag{6}$$

Where $\mathbf{b}(n) \in R^{n_2 \times n_2}$ and $\mathbf{b}(q) \in R^{n_2 \times n_2}$ represent size $n_2 \times n_2$ image patches centered at pixel $\mathbf{u}(n)$ and $\mathbf{u}(q)$, respectively; parameter $h$ controls the attenuation speed of weight function.

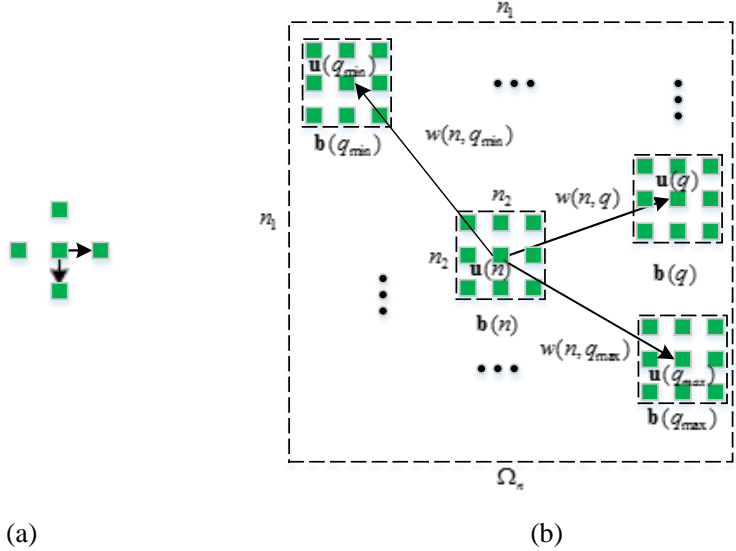

(a)            (b)

**Figure 2** Gradient and graph derivatives.(a) Gradient; (b) graph derivatives.

In this study, the concept of graph derivative is extended to multi-energy image $\mathbf{u} = [\mathbf{u}_1^T, \cdots, \mathbf{u}_m^T, \cdots \mathbf{u}_M^T]^T \in R^{N_1 N_2 M \times 1}$, and nonlocal Jacobian matrix $J_{NL,\mathbf{w}_n}\mathbf{u}(n) \in R^{M \times Q}$ is defined as

$$J_{NL,\mathbf{w}_n}\mathbf{u}(n) = \begin{bmatrix} \nabla_{\mathbf{w}_n}\mathbf{u}_1(n) \\ \vdots \\ \nabla_{\mathbf{w}_n}\mathbf{u}_m(n) \\ \vdots \\ \nabla_{\mathbf{w}_n}\mathbf{u}_M(n) \end{bmatrix} = \begin{bmatrix} \nabla_{w,q_{\min}}\mathbf{u}_1(n) & \cdots & \nabla_{w,q}\mathbf{u}_1(n) & \cdots & \nabla_{w,q_{\max}}\mathbf{u}_1(n) \\ \vdots & & \vdots & & \vdots \\ \nabla_{w,q_{\min}}\mathbf{u}_m(n) & \cdots & \nabla_{w,q}\mathbf{u}_m(n) & \cdots & \nabla_{w,q_{\max}}\mathbf{u}_m(n) \\ \vdots & & \vdots & & \vdots \\ \nabla_{w,q_{\min}}\mathbf{u}_M(n) & \cdots & \nabla_{w,q}\mathbf{u}_M(n) & \cdots & \nabla_{w,q_{\max}}\mathbf{u}_M(n) \end{bmatrix} \tag{7}$$

Where $Q$ is the total number of pixels in $\Omega_n$ and $\mathbf{w}_n = [w(n, q_{\min}), \cdots, w(n,q), \cdots w(n, q_{\max})]$.

In order to save computation and reduce the influence of weights on the measurement of image structure similarity at different energy spectrum, the weights for pixel $\mathbf{u}(n)$ are modified from equation (6) as followings:

$$w(n,q) = \begin{cases} 1, q \in S, q \in \Omega_n \\ 0, q \notin S, q \in \Omega_n \end{cases} \qquad (8)$$

Where $S$ is defined as the index set of the central pixels of the $L$ image patches which are the most similar ( the first $L$ maximum weights according to equation (6) ) to the current image patch (the image patch centered at the pixel $\mathbf{u}(n)$) in the search window $\Omega_n$. Obviously, only when $w(n,q)=1$, the pixel pairs are connected and the derivative operation is performed. By using weight $w(n,q)$ to define the connectivity between pairs of pixels $(n,q)$, we can flexibly describe the spatial nonlocal similarity of the image. In order to better couple the structural similarity of images at different energy spectrum, the weights at different energy spectrum are unified into the same distribution. Hence, the size of $J_{NL,\mathbf{w}_n}\mathbf{u}(n)$ reduces from $M \times Q$ to $M \times L$, which reduces the computation burden further.

As the reconstructed images at different energy spectrum have similar structures with each other, the graph derivatives of the images are approximately parallel, $J_{NL,\mathbf{w}_n}\mathbf{u}(n)$ has the low rank structure, which is equivalent to the minimization of nuclear norm in mathematical sense. Hence, we define the NLTNV as follows:

$$NLTNV(\mathbf{u}) = \sum_{n=1}^{N} \| J_{NL,\mathbf{w}_n}\mathbf{u}(n) \|_* = \sum_{n=1}^{N} \sum_{t=1}^{Q} \sigma_{n,t} \qquad (9)$$

Where $\sigma_{n,t}$ is the $t^{th}$ singular value of $J_{NL,\mathbf{w}_n}\mathbf{u}(n)$.

Although NLTNV can effectively use nonlocal prior information to restore the structural features of the image, it still only uses the first derivative information of the image, which tends to lead to the stepwise effect, especially for the smooth part of the reconstructed image. Several works [14]-[16] have demonstrated incorporating the higher derivative term can alleviate this problem effectively. Hence, we introduce higher order derivatives to define structure similarity. First, we define the matrix $J^2 U(i,j)$ with the second order derivatives as follows:

$$J^2\mathbf{U}(i,j) = \begin{bmatrix} \nabla_{xx}^2 \mathbf{U}_1(i,j) & \nabla_{xy}^2 \mathbf{U}_1(i,j) & \nabla_{yx}^2 \mathbf{U}_1(i,j) & \nabla_{yy}^2 \mathbf{U}_1(i,j) \\ \vdots & \vdots & \vdots & \vdots \\ \nabla_{xx}^2 \mathbf{U}_m(i,j) & \nabla_{xy}^2 \mathbf{U}_m(i,j) & \nabla_{yx}^2 \mathbf{U}_m(i,j) & \nabla_{yy}^2 \mathbf{U}_m(i,j) \\ \vdots & \vdots & \vdots & \vdots \\ \nabla_{xx}^2 \mathbf{U}_M(i,j) & \nabla_{xy}^2 \mathbf{U}_M(i,j) & \nabla_{yx}^2 \mathbf{U}_M(i,j) & \nabla_{yy}^2 \mathbf{U}_M(i,j) \end{bmatrix} \qquad (10)$$

Then, we define the total nuclear second order variation as

$$TNSV(\mathbf{U}) = \sum_{i=1}^{N_1} \sum_{j=1}^{N_2} \| J^2\mathbf{U}(i,j) \|_* \qquad (11)$$

The TNSV can alleviate the stepwise effect of NLTNV and promote the parallel of the second order derivative of the image at different energy spectrum, which is beneficial for the piece-wise smoothing parts of the reconstructed image.

Therefore, using the above definitions with respect to NLTNV and TNSV, the NLTNGV regularization term is defined as follows:

$$NLTNGV(\mathbf{u}) = \alpha_1 NLTNV(\mathbf{u}) + \alpha_2 TNSV(\mathbf{U})$$
$$= \alpha_1 \sum_{n=1}^{N} \| J_{NL,\mathbf{w}_n} \mathbf{u}(n) \|_* + \alpha_2 \sum_{i=1}^{N_1} \sum_{j=1}^{N_2} \| J^2 \mathbf{U}(i,j) \|_* \quad (12)$$

Where $\alpha_1, \alpha_2 \geq 0$ are the balance coefficients for NLTNV and TNSV.

In order to show that NLTNGV regularization can promote the sparsity of nonlocal gradient and local second order derivative, the spatial nonlocal self similarity and the structural similarity of the reconstructed images at different energy spectrum simultaneously, the following brief analysis is given. First, the F norm for a general matrix is introduced: for matrix $\mathbf{C} \in R^{m \times n}$, the F norm is defined as $\| \mathbf{C} \|_F = (\sum_{i=1}^{m} \sum_{j=1}^{n} |c_{i,j}|^2)^{\frac{1}{2}}$. According to the definition of F norm, we can get

$$\| J_{NL,\mathbf{w}_n} \mathbf{u}(n) \|_F = \sqrt{\sum_{m=1}^{M} \sum_{q=q_{min}}^{q_{max}} | \nabla_{w,q} \mathbf{u}_m(n) |^2} \quad (13)$$

$$\| J^2 \mathbf{U}(i,j) \|_F = \sqrt{\sum_{m=1}^{M} (| \nabla_{xx}^2 \mathbf{U}_m(i,j) |^2 + | \nabla_{xy}^2 \mathbf{U}_m(i,j) |^2 + | \nabla_{yx}^2 \mathbf{U}_m(i,j) |^2 + | \nabla_{yy}^2 \mathbf{U}_m(i,j) |^2)} \quad (14)$$

Furthermore, we introduce the theorem: if $\sigma_1 \geq \sigma_2 \geq \cdots \geq \sigma_r \geq 0$ is singular value of matrix $\mathbf{C} \in R^{m \times n}$, then $\| \mathbf{C} \|_F = \sqrt{\sigma_1^2 + \sigma_2^2 + \cdots + \sigma_r^2}$. By applying the above theorem to nonlocal Jacobian matrix $J_{NL} \mathbf{u}(n)$, we have

$$\| J_{NL,\mathbf{w}_n} \mathbf{u}(n) \|_F^2 = \sum_{t=1}^{Q} \sigma_{n,t}^2 \quad (15)$$

For $\sigma_{n,1} \geq \cdots \geq \sigma_{n,t} \geq \cdots \geq \sigma_{n,Q} \geq 0$, the following inequality holds:

$$\sum_{t=1}^{Q} \sigma_{n,t}^2 \leq (\sum_{t=1}^{Q} \sigma_{n,t})^2 \quad (16)$$

According to the above inequality, also combined with the definition of nuclear norm, we have

$$\| J_{NL,\mathbf{w}_n} \mathbf{u}(n) \|_F \leq \| J_{NL,\mathbf{w}_n} \mathbf{u}(n) \|_* \quad (17)$$

Furthermore, consider the above equality for the all pixels, then

$$\sum_{n=1}^{N} \| J_{NL,\mathbf{w}_n} \mathbf{u}(n) \|_F \leq \sum_{n=1}^{N} \| J_{NL,\mathbf{w}_n} \mathbf{u}(n) \|_* \quad (18)$$

In a similar way, we can derive the following relationship:

$$\sum_{i=1}^{N_1}\sum_{j=1}^{N_2}\|J^2\mathbf{U}(i,j)\|_F \leq \sum_{i=1}^{N_1}\sum_{j=1}^{N_2}\|J^2\mathbf{U}(i,j)\|_* \tag{19}$$

Combining (18) and (19), we obtain

$$\alpha_1\sum_{n=1}^{N}\|J_{NL,\mathbf{w}_n}\mathbf{u}(n)\|_F + \alpha_2\sum_{i=1}^{N_1}\sum_{j=1}^{N_2}\|J^2\mathbf{U}(i,j)\|_F \leq NLTNGV(\mathbf{u}) \tag{20}$$

From the perspective of optimization, minimizing NLTNGV can reduce both the value of $\sum_{n=1}^{N}\|J_{NL,\mathbf{w}_n}\mathbf{u}(n)\|_F$ and $\sum_{i=1}^{N_1}\sum_{j=1}^{N_2}\|J^2\mathbf{U}(i,j)\|_F$. When NLTNGV reaches a small value, both $\sum_{n=1}^{N}\|J_{NL,\mathbf{w}_n}\mathbf{u}(n)\|_F$ and $\sum_{i=1}^{N_1}\sum_{j=1}^{N_2}\|J^2\mathbf{U}(i,j)\|_F$ take a small value as well. It means that NLTNGV regularization can promote the sparsity of nonlocal gradients and local second derivatives of the image at each energy spectrum, respectively. In other words, NLTNGV can effectively exploit the sparse prior information.

On the other hand, the nuclear norm is the most compact convex relaxation of matrix rank, minimizing NLTNGV will promote $J_{NL,\mathbf{w}_n}\mathbf{u}(n)$ and $J^2\mathbf{U}(i,j)$ to have low rank property, thus promote the graph derivative and second-order derivative of image at each energy spectrum to share common direction, respectively. In other words, NLTNGV can promote the images to share structure similarity at each energy spectrum.

As the NLTNGV regularization needs to calculate the graph derivative according to the non local self similarity of the image, the initial image $\mathbf{u}_{ini}$ needs to be reconstructed to calculate the above weights. For voltage switching scanning manner studied in this work, we particularly consider the three energy case ($M=3$). As the projection information at different energy spectrum is complementary, $\mathbf{u}_{ini}$ can be reconstructed by fusing the projection information at three energy spectrum. In this work we use notions $\mathbf{P_1},\mathbf{P_2},\mathbf{P_3}\in R^{n_d\times S}$ ($n_d$ is the number of detector elements, $S$ is the total number of sampling views) to represent the projection data arranged according to the order of the sampling view (sinogram) at the low, medium and high energy spectrum respectively.

In order to minimize the influence of the numerical difference of projection data at different energy spectrums on the reconstruction results of $\mathbf{u}_{ini}$, the projection of high-energy spectrum are taken as the reference, and the projection data of the other two energy spectrums are normalized into the high-energy projection $\mathbf{P_{1\to 3}}$ and $\mathbf{P_{2\to 3}}$, namely $\mathbf{P_{1\to 3}}=\eta_1\mathbf{P_1}$, $\mathbf{P_{2\to 3}}=\eta_2\mathbf{P_2}$, where $\eta_1$ and $\eta_2$ are the normalization coefficients. After normalization, the fused sinogram is obtained as follows:

$$\begin{aligned}\mathbf{P} &= [\mathbf{P}_{1\to 3}^1,\mathbf{P}_{2\to 3}^1,\mathbf{P}_3^1,\cdots,\mathbf{P}_{1\to 3}^s,\mathbf{P}_{2\to 3}^s,\mathbf{P}_3^s\cdots,\mathbf{P}_{1\to 3}^S,\mathbf{P}_{2\to 3}^S,\mathbf{P}_3^S] \\ &= [\eta_1\mathbf{P}_1^1,\eta_2\mathbf{P}_2^1,\mathbf{P}_3^1,\cdots,\eta_1\mathbf{P}_1^s,\eta_2\mathbf{P}_2^s,\mathbf{P}_3^s\cdots,\eta_1\mathbf{P}_1^S,\eta_2\mathbf{P}_2^S,\mathbf{P}_3^S]\end{aligned} \tag{21}$$

Where $\mathbf{P}_m^s \in R^{n_d \times 1}$ is the projection of the $s^{th}$ sampling view at the $m^{th}$ energy spectrum.

In order to determine $\eta_1$ and $\eta_2$, we simply use projection operation, namely

$$\eta_1 = \frac{\mathbf{p}_1^T \mathbf{p}_3}{\|\mathbf{p}_1\|_2^2}$$
$$\eta_2 = \frac{\mathbf{p}_2^T \mathbf{p}_3}{\|\mathbf{p}_2\|_2^2} \tag{22}$$

Where $\mathbf{p}_1, \mathbf{p}_2, \mathbf{p}_3$ are the vectorization of $\mathbf{P}_1, \mathbf{P}_2, \mathbf{P}_3$. After obtaining the projection $\mathbf{P}$, we use the TV based regularization method [17] to reconstruct the initial image $\mathbf{u}_{ini}$.

### 2.3 Multi-energy CT reconstruction using NLTNGV

Using the NLTNGV regularization defined above, we propose the following model for multi-energy CT reconstruction:

$$NLTNGV(\mathbf{u}) + \frac{\mu}{2} \| A\mathbf{u} - \mathbf{p} \|^2 \tag{23}$$

By using the definition of NLTNGV, the above model can be rewritten as

$$\alpha_1 \sum_{n=1}^{N} \| J_{NL,\mathbf{w}_n} \mathbf{u}(n) \|_* + \alpha_2 \sum_{i=1}^{N_1} \sum_{j=1}^{N_2} \| \mathbf{J}^2 \mathbf{U}(i,j) \|_* + \frac{\mu}{2} \| A\mathbf{u} - \mathbf{p} \|^2 \tag{24}$$

Furthermore, by introducing variables $\mathbf{X}_n$ and $\mathbf{Y}_{i,j}$, (24) is equivalent to the following constrained optimization problem:

$$\min_{u} \alpha_1 \sum_{n=1}^{N} \| \mathbf{X}_n \|_* + \alpha_2 \sum_{i=1}^{N_1} \sum_{j=1}^{N_2} \| \mathbf{Y}_{i,j} \|_* + \frac{\mu}{2} \| A\mathbf{u} - \mathbf{p} \|^2$$
$$s.t. J_{NL,\mathbf{w}_n} \mathbf{u}(n) = \mathbf{X}_n, n = 1, 2, \cdots N \tag{25}$$
$$J^2 \mathbf{U}(i,j) = \mathbf{Y}_{i,j}, i = 1, \cdots N_1, j = 1, \cdots N_2$$

The corresponding augmented Lagrange function is given by

$$L(\mathbf{u}, \mathbf{X}, \mathbf{Y}, \lambda, \gamma) = \alpha_1 \sum_{n=1}^{N} \| \mathbf{X}_n \|_* + \alpha_2 \sum_{i=1}^{N_1} \sum_{j=1}^{N_2} \| \mathbf{Y}_{i,j} \|_* + \frac{\mu}{2} \| A\mathbf{u} - \mathbf{p} \|^2 +$$
$$\frac{\rho}{2} \sum_{n=1}^{N} \| J_{NL,\mathbf{w}_n} \mathbf{u}(n) - \mathbf{X}_n + \lambda_n \|^2 + \frac{\rho}{2} \sum_{i=1}^{N_1} \sum_{j=1}^{N_2} \| J^2 \mathbf{U}(i,j) - \mathbf{Y}_{i,j} + \gamma_{i,j} \|^2 \tag{26}$$

By using ADMM algorithm, the optimization of (26) can be converted into alternatively solving the following sub-problems:

$$\begin{cases} \mathbf{X}^{k+1} = \arg\min_{\mathbf{X}} L(\mathbf{X}, \mathbf{Y}^k, \mathbf{u}^k, \lambda^k, \gamma^k) \\ \mathbf{Y}^{k+1} = \arg\min_{\mathbf{Y}} L(\mathbf{X}^k, \mathbf{Y}, \mathbf{u}^k, \lambda^k, \gamma^k) \\ \mathbf{u}^{k+1} = \arg\min_{u} L(\mathbf{X}^{k+1}, \mathbf{Y}^{k+1}, \mathbf{u}, \lambda^k, \gamma^k) \\ \lambda_n^{k+1} = \lambda_n^k + (J_{NL,\mathbf{w}_n} \mathbf{u}^{k+1}(n) - \mathbf{X}_n^{k+1}), n = 1, 2, \ldots N \\ \gamma_{i,j}^{k+1} = \gamma_{i,j}^k + J^2 \mathbf{U}^{k+1}(i,j) - \mathbf{Y}_{i,j}^{k+1}, i = 1, 2, \cdots N_1, j = 1, 2, \cdots N_2 \end{cases} \tag{27}$$

**(1) Solving $\mathbf{X}, \mathbf{Y}$ subproblems**

The $\mathbf{X}, \mathbf{Y}$ subproblems are separable and can be further decomposed into the sub-problems

as follows:

$$\mathbf{X}_n^{k+1} = \arg\min_{\mathbf{X}_n} \alpha_1 \|\mathbf{X}_n\|_* + \frac{\rho}{2} \| J_{NL,\mathbf{w}_n} \mathbf{u}^{k+1}(n) - \mathbf{X}_n + \lambda_n^k \|^2, n=1,2,\cdots N \tag{28}$$

$$\mathbf{Y}_{i,j}^{k+1} = \arg\min_{\mathbf{Y}_{i,j}} \alpha_2 \|\mathbf{Y}_{i,j}\|_* + \frac{\rho}{2} \| \mathbf{J}^2 \mathbf{U}^k(i,j) - \mathbf{Y}_{i,j} + \gamma_{i,j}^k \|^2, i=1,2,\cdots N_1, j=1,2,\cdots N_2 \tag{29}$$

The minimization problem (28),(29) are the proximal problems with respect to nuclear norm, which can be solved efficiently. According to [18], the analytic solutions are expressed as

$$\mathbf{X}_n^{k+1} = SVT_{\alpha_1/\rho}(J_{NL,\mathbf{w}_n}\mathbf{u}^k(n) + \lambda_n^k), n=1,2,\cdots N \tag{30}$$

$$\mathbf{Y}_{i,j}^{k+1} = SVT_{\alpha_2/\rho}(\mathbf{J}^2\mathbf{U}^k(i,j) + \gamma_{i,j}^k), i=1,\cdots N_1, j=1,\cdots N_2 \tag{31}$$

Where $SVT_\tau(\cdot)$ is the singular value threshold operation, given by

$$SVT_\tau(\mathbf{X}) = \mathbf{V}_1^T diag\left(\max(\sigma_t - \tau, 0)\right)\mathbf{V}_2 \tag{32}$$

Where $\mathbf{X} = \mathbf{V}_1^T \Sigma \mathbf{V}_2$ is the singular value decomposition of the matrix $\mathbf{X}$, and $\sigma_t = \Sigma(t,t)$ is the $t^{th}$ singular value, the notation $diag()$ represents a diagonal matrix.

**(2) Solving u subproblems**

With $\mathbf{X}_n, \mathbf{Y}_{i,j}, \lambda_n, \gamma_{i,j}$ fixed as $\mathbf{X}_n^{k+1}, \mathbf{Y}_{i,j}^{k+1}, \lambda_n^k, \gamma_{i,j}^k$, the subproblem $\mathbf{u}$ is a convex quadratic optimization problem, given by

$$\mathbf{u}^{k+1} = \arg\min_u \frac{\mu}{2}\|A\mathbf{u}-\mathbf{p}\|^2 + \frac{\rho}{2}\sum_{n=1}^{N}\|J_{NL,\mathbf{w}_n}\mathbf{u}(n) - \mathbf{X}_n^{k+1} + \lambda_n^k\|^2 + \frac{\rho}{2}\sum_{i=1}^{N_1}\sum_{j=1}^{N_2}\|\mathbf{J}^2\mathbf{U}(i,j) - \mathbf{Y}_{i,j}^{k+1} + \gamma_{i,j}^k\|^2 \tag{33}$$

By using the first-order optimality condition that the gradient of right hand of (33) should be 0, we can get

$$\mu A^T(A\mathbf{u}-\mathbf{p}) + \rho\sum_{n=1}^{N} J_{NL,\mathbf{w}_n}^T(\mathbf{u}(n) - \mathbf{X}_n^k + \lambda_n^k) + \rho\sum_{i=1}^{N_1}\sum_{j=1}^{N_2} J^{2T}(\mathbf{J}^2\mathbf{U}(i,j) - \mathbf{Y}_{i,j}^k + \gamma_{i,j}^k) = 0 \tag{34}$$

By some simple manipulation, the following system of linear equations are obtained:

$$(\mu A^T A + \rho\sum_{n=1}^{N} J_{NL,\mathbf{w}_n}^T J_{NL,\mathbf{w}_n} + \rho J^{2T} J^2)\mathbf{u} = \mu A^T \mathbf{p} + \rho\sum_{n=1}^{N} J_{NL,\mathbf{w}_n}^T(\mathbf{X}_n^{k+1} - \lambda_n^k) + \rho\sum_{i=1}^{N_1}\sum_{j=1}^{N_2} J^{2T}(\mathbf{Y}_{i,j}^{k+1} - \gamma_{i,j}^k) \tag{35}$$

As the system matrix of (35) is symmetric positive, the conjugate gradient method [19] can be used to solve it efficiently.

**(3) Update multipliers**

$$\begin{cases} \lambda_n^{k+1} = \lambda_n^k + (J_{NL,\mathbf{w}_n}\mathbf{u}^{k+1}(n) - \mathbf{X}_n^{k+1}), n=1,2,\ldots N \\ \gamma_{i,j}^{k+1} = \gamma_{i,j}^k + \mathbf{J}^2\mathbf{U}^{k+1}(i,j) - \mathbf{Y}_{i,j}^{k+1}, i=1,2,\cdots N_1, j=1,2,\cdots N_2 \end{cases} \tag{36}$$

The implementation steps of the proposed algorithm for multi-energy CT image reconstruction are summarized as algorithm 1. The convergence of the proposed algorithm is the same as that of the classical ADMM algorithm given by [20].

<div align="center">Algorithm 1</div>

**Multi-energy CT reconstruction using NLTNGV regularization**

**Initialization:** Use $\mathbf{u}_{ini}$ to calculate $\{w(n,q)\}$, set tolerance value $\varepsilon$ and maximum iteration number $N_{\max}$, $k=0$, $\mathbf{u}^0 = 0$.

**Start:**

**While** ( $k \leq N_{\max}$ and $\|\mathbf{u}^{k+1} - \mathbf{u}^k\| > \varepsilon$ )

(1) Update $\mathbf{X}_n^{k+1}$ according to equation (30);

　　Update $\mathbf{Y}_{i,j}^{k+1}$ according to equation (31);

(2) Use conjugate gradient algorithm to solve (35) to obtain $\mathbf{u}^{k+1}$;

(3) Update multiplier:

　　Update $\boldsymbol{\lambda}_n^{k+1}, \boldsymbol{\gamma}_{i,j}^{k+1}$ according to equation (36);

　　$k = k+1$

**End**

**Output:** reconstructed image $\mathbf{u}$.

## 3. Experiment and results

In order to verify the validity of the proposed regularization, real data experiments are performed in this section. The experimental results are compared with those of the method based on TNV regularization [4]. For convenience, the reconstruction based on the proposed NLTNV regularization (equation (12) with $\alpha_1 \neq 0, \alpha_2 = 0$) is called NLTNV method, while the reconstruction based on proposed NLTNGV regularization (equation (12) with $\alpha_1 \neq 0, \alpha_2 \neq 0$) is called NLTNGV method. The fan-beam scanning is adopted in the experiments, and the corresponding imaging parameters are shown in Table 3.1.

<div align="center">Table 3.1</div>

| CT geometric imaging parameters |
| --- |
| Distance from X-ray source to detector: 1400mm |
| Distance from X-ray source to rotation axis: 1000mm |
| Detector resolution: 1024 |
| Element size of detector: 0.2mm |
| Size of reconstructed image: $256 \times 256$ |
| Pixel size : $0.5\text{mm} \times 0.5\text{mm}$ |

In this work, PSNR, RMSE and SSIM [21] are used to evaluate the quality of the reconstructed images quantitatively, the formulas of which are given by

$$PSNR = 10\log_{10}\frac{(\max(\mathbf{u}_{m,ref}))^2}{\sum_{n=1}^{N_1 \times N_2}\frac{1}{N_1 N_2}(\mathbf{u}_m(n)-\mathbf{u}_{m,ref}(n))^2} \quad (37)$$

$$RMSE = \frac{1}{\sqrt{N_1 N_2}}(\sum_{n=1}^{N_1 \times N_2}(\mathbf{u}_m(n)-\mathbf{u}_{m,ref}(n)))^{1/2} \quad (38)$$

$$SSIM = \frac{2\bar{u}_{m,ref}\bar{u}_m(2v_{12}+c_2)}{(\bar{u}_{m,ref}^2+\bar{u}_m^2+c_1)(v_1^2+v_2^2+c_2)} \quad (39)$$

Where $\bar{u}_m$ and $\bar{u}_{m,ref}$, $v_1$, $v_2$ and $v_{12}$ are the mean, variance and covariance of the reconstructed image $\mathbf{u}_m$ and the reference image $\mathbf{u}_{m,ref}$ of the $m^{th}$ energy spectrum, respectively, $c_1$ and $c_2$ are constants for numerical stability. As the above metrics are defined for two-dimensional image at the $m^{th}$ spectrum, the mean values of all the M spectrums are taken as the numerical evaluation.

**3.1 Parameter setting**

The proposed algorithm is based on the graph derivative to realize the regularization, so it is necessary to determine the size of the region (search window $\Omega$) in which each pixel searches for pixels that is most similar to it and the size of the image patch to calculate the nonlocal self similarity weights. In order to balance the computation burden and performance, the search window is set to 11pixel×11pixel, the image patch size is set to 3 pixel×3 pixel and the number of most similar pixels for each pixel is selected as $L=8$. In order to fairly compare the performance of the competing methods, the optimized regularization parameters are selected for each method in terms of obtaining best RMSE.

**3.2 Real data experiment**

In the experiment, the X-ray source voltage is set to 80kVp, 110kVp and 140kVp respectively, and the current is set to 0.2mA to scan a nut sample. At 80kvp, 60 view projection data are evenly collected in $[0,360^0)$ with the sampling view starting at $0^0$, and the projection data of the other two voltages are collected in the same way except that the sampling angle starts at $2^0$ and $4°$, respectively. All the above projection data are used for TNV, NLTNV and NLTNGV for image reconstruction.

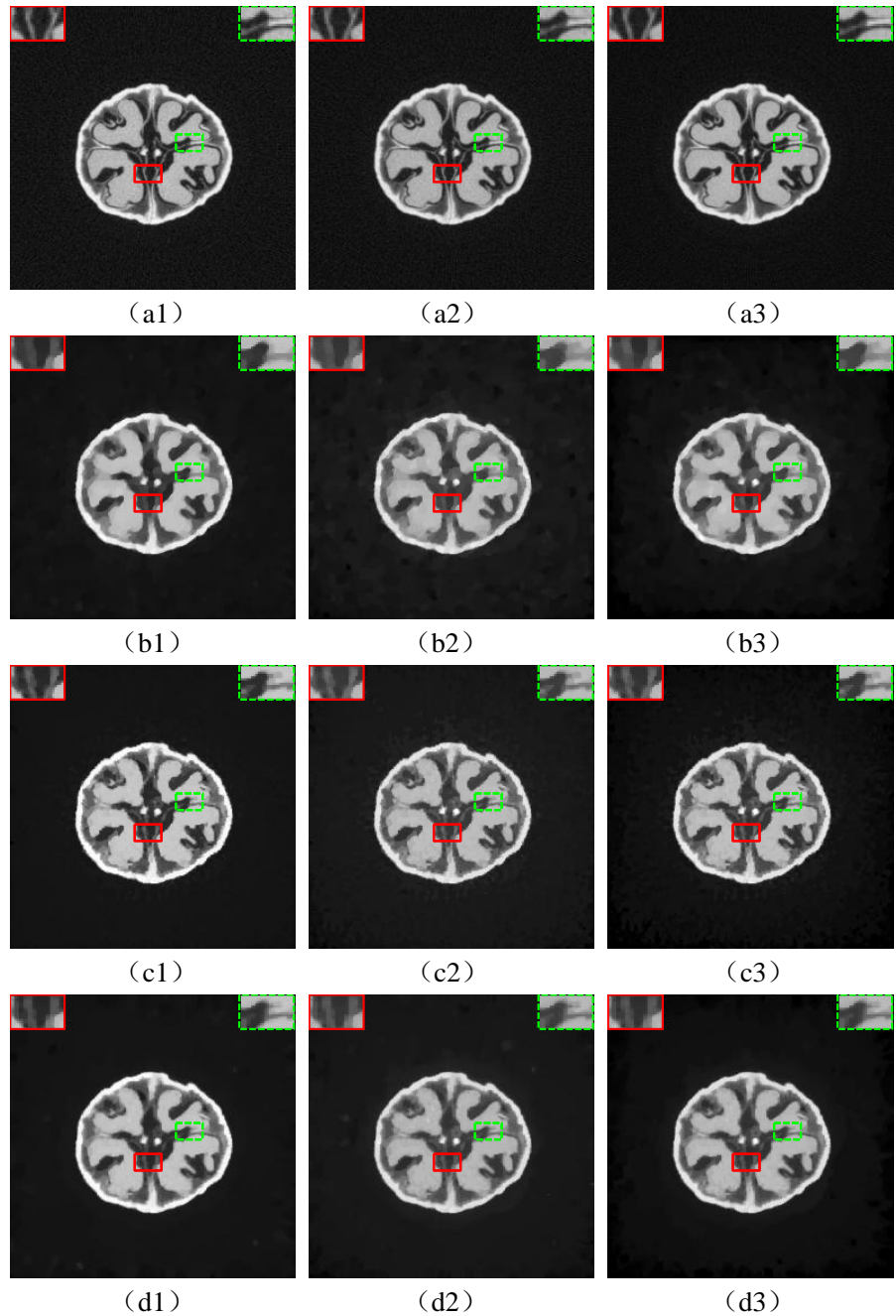

**Figure 3** Reconstruction results of different methods. (a) Reference images; (b) images reconstructed by TNV method; (c) images reconstructed by NLTNV method; (d) images reconstructed by NLTNGV method. The first column shows the 80kVp reconstruction results, the second column shows the 110kVp reconstruction results, and the third column shows the 140kVp reconstruction results.

Figure 3 (a1) - (a3) are the reference images reconstructed by FBP algorithm from high dose projection data; Figure 3 (b1) - (b3) are the images reconstructed by TNV method; Figure 3 (c1) - (c3) are the images reconstructed by NLTNV method; Figure 3 (d1) - (d3) are the images reconstructed by NLTNGV method. From the reconstruction results, we can see that all the methods can effectively eliminate the artifacts caused by sparse sampling, but they have different performance in restoring the details of the images in noisy case. Figure 3 (b1) - (b3) show that although the TNV method can suppress noise effectively, the reconstructed images are over smoothed. This is because TNV only uses the very local neighborhood information of the pixel to couple the structural similarity information along energy dimension, thus inevitably smoothes out the details of the image while suppressing the noise. From the magnified parts, it can be observed that the details with low contrast are blurred and difficult to be distinguished. In contrast, the details of the reconstructed images in Figure 3(c1)-(c3) are much clearer, meanwhile noise is well suppressed. This is due to the fact that NLTNV method can take good advantage of the spatial nonlocal self similarity prior information to restore the structural features. However, some stepwise effect can be still observed. Compared with NLTNV method, due to the introduced second derivative regularization term in NLTNGV method, the stepwise effect in Figure 3(d1)-(d3) are much reduced and more visually natural. The RMSE, PSNR and SSIM of each reconstruction method are given in table 3.2. It can be seen from the quantitative evaluation that the NLTNGV method performs best. The experimental results show that the overall quality of the reconstructed images using the proposed regularization is better than that of the TNV regularization based method.

Table 3.2 Quantitative evaluation results

|         | RMSE     | PSNR  | SSIM   |
|---------|----------|-------|--------|
| TNV     | 0.001089 | 26.08 | 0.7291 |
| NLTNV   | 0.001019 | 26.62 | 0.7381 |
| NLTNGV  | 0.001018 | 26.66 | 0.7417 |

**3. Discussion and conclusion**

In this paper, we use the low rank property of the first and second derivative of multi-energy CT images to construct the structural similarity measure, and propose the NLTNGV regularization term. NLTNGV is composed of NLTNV and TNSV. The NLTNV is based on the nonlocal first derivative information of the image and can exploit a variety of prior information, but it is not very favorable to preserve the smooth transition of the image. By further introducing the second derivative term, this disadvantage can be overcome effectively. The primary experimental results show that NLTNGV-based reconstruction method has the significant advantage over the TNV-based reconstruction method by introducing the nonlocal self similarity prior information and second order derivative term, and can greatly improve the overall quality of the reconstructed image. The NLTNGV regularization method can also be applied to other multi-energy CT image reconstruction problems, such as projection data acquired from photon counting detector, where the initial image should be replaced by the full spectrum image. Meanwhile, the proposed method can be readily extended for material decomposition in image domain. Furthermore, the TNSV regularization can also be extended to nonlocal version for further improvement. In addition, the NLTNGV regularization can be extended as tensor form or tensor decomposition for nonlocal similar patches in gradient domain and this work is underway.